\documentclass{emulateapj}

\newcommand{\tdm}{\ensuremath{\Delta t_{\rm DM}}}
\newcommand{\tobs}{\ensuremath{\Delta t_{\rm obs}}}
\newcommand{\dm}{\ensuremath{{\rm pc}\,{\rm cm}^{-3}}}
\newcommand{\frb}{FRB~150418}
\newcommand{\omegaigm}{\ensuremath{\Omega_{\rm IGM}}}
\slugcomment{\today}

\begin{document}

\title{Limits on Einstein's Equivalence Principle from the first localized Fast Radio Burst FRB 150418}

\author{S.~J. Tingay\altaffilmark{1} and D.~L. Kaplan\altaffilmark{2}}

\altaffiltext{1}{International Centre for Radio Astronomy Research
  (ICRAR), Curtin University, Bentley, WA 6102, Australia; \email{stingay@curtin.edu.au}}
\altaffiltext{2}{Department of Physics, University of
  Wisconsin--Milwaukee, Milwaukee, WI 53201, USA; \email{kaplan@uwm.edu}}

\begin{abstract}
Fast Radio Bursts have recently been used to place limits on
Einstein's Equivalence Principle via observations of time delays
between photons of different radio frequencies by \citet{wei15}.
These limits on differential post-Newtonian parameters ($\Delta
\gamma<2.52\times10^{-8}$) are the best yet achieved but still rely on
uncertain assumptions, namely the relative contributions of dispersion
and gravitational delays to the observed time delays and the distances
to FRBs.  Also very recently, the first FRB host galaxy has likely been
identified, providing the first redshift-based distance estimate to
FRB 150418 \citep{kea16}.  Moreover, consistency between the
\omegaigm\ estimate from FRB 150418 and \omegaigm~expected from
$\Lambda$CDM models and WMAP observations leads one to conclude that
the observed time delay for FRB 150418 is highly dominated by
dispersion, with any gravitational delays small contributors.  This
points to even tighter limits on $\Delta \gamma$.  In this paper, the
technique of \citet{wei15} is applied to FRB 150418 to produce a limit
of $\Delta \gamma < 1 - 2\times10^{-9}$, approximately an order of
magnitude better than previous limits and in line with expectations by
\citet{wei15} for what could be achieved if the dispersive delay is
separated from other effects.  Future substantial improvements in such
limits will depend on accurately determining the contribution of
individual ionized components to the total observed time delays for
FRBs.
\end{abstract} 


\section{INTRODUCTION}
Fast Radio Bursts (FRBs), discovered by \cite{lor07} and reported by a
range of teams since
\citep{kea16,cha15,mas15,pet15,rav15,bur14,spi14,tho13,kea12} have been
predicted to be fantastic cosmological and astrophysical probes
\citep[e.g.,][]{mac15}.  They have additionally been shown by
\citet{wei15} to be an excellent tool for constraining fundamental
physics, namely Einstein's
Equivalence Principle (EEP), if FRBs originate at cosmological
distances.  \citet{wei15} use limits on the gravitational time delay
between photons of different radio frequencies generated by a FRB to
infer upper limits on $\Delta \gamma=\gamma_{1}-\gamma_{2}$, where
$\gamma_{n}$ is the post-Newtonian parameter describing how much space
curvature is produced by a unit test mass \citep{mis73} at the
particular frequency of photon $n$; $\gamma=1$ for General Relativity.

Effectively, this treatment looks for whether or not photons of
different energies ``fall'' at the same rate in a gravitational
potential\footnote{This test is
similar to that performed on solar neutrinos
\citep{gas88,hal91,gag00}, where the violation is a function of
neutrino flavor rather than energy.}, in contrast to absolute limits on $\gamma-1$ that must be
calibrated more carefully \citep{wil14}, with the best
limits around $10^{-4}$ or $10^{-5}$ \citep{lam11,ber03}.  \citet{wei15} infer their
limits from their equations 3 and 4:

\[\tobs - \tdm > \frac{\gamma_{1}-\gamma_{2}}{c^{3}}\int^{r_{e}}_{r_{o}}U(r)dr ,\]

\[\Delta \gamma = \gamma_{1}-\gamma_{2} < (\tobs - \tdm
)\left[\frac{GM_{\rm MW}}{c^{3}}\ln\left(\frac{d}{b}\right)\right]^{-1} ,\]

where: \tobs\ is the observed time delay between photons of different
frequencies; \tdm\ is the time delay between photons of different
frequencies due only to plasma dispersion, important at radio
frequencies; $\gamma$ is as described above; $c$ is the speed of
light; $r_{e}$ and $r_{o}$ are the positions of the emitter and the
observer, respectively; $U(r)$ is the gravitational potential along
the line of sight from emitter to observer; $G$ is the gravitational
constant; $M_{\rm MW}$ is the mass of the Milky Way; $d$ is the
distance between emitter and observer; and $b$ is the impact parameter
of the emitted signal relative to the Milky Way center. Note that more
accurate potential models can be used (e.g., \citealt{wu16}) but they
change the results by at most 10\%.  \citet{wei15} approximate the
time delay between photons of two different frequencies due to the
gravitational potential along the line of sight as $\Delta t_{\rm
  grav}=\tobs - \tdm$.

The fundamental observable for FRBs is \tobs.  For all detected FRBs
the full value of \tobs\ has been equated with \tdm\ to infer a
dispersion measure (DM) and therefore estimate a distance for each
FRB, assuming knowledge of the ionized intergalactic medium and the
FRB's local environment.  This
indirect method of estimating distances has been necessary as the
single dish radio telescopes used to detect FRBs have lacked the
angular resolution to identify any progenitor or host galaxy.

\citet{wei15} use such a line of reasoning (that $\tobs\sim\tdm$) to
estimate the redshift, and therefore distance, to FRB 110220
\citep{tho13}, but then require that the majority of \tobs\
is due to gravitational delay ($\tobs \gg \tdm$) in order to produce a
strict upper limit of $\Delta \gamma<2.52\times10^{-8}$ between the
frequencies of 1.2\,GHz and 1.5\,GHz.  \citet{wei15} undertake a
sensitivity analysis, given the uncertainties in the assumptions
(distance to FRB and relative contributions of dispersion delay and
gravitational delay to \tobs), to show that their limits are
not particularly sensitive to the assumptions.

\citet{wei15} also produced limits for two proposed gamma-ray burst (GRB)/FRB
associations.  GRB 101011A and GRB 100704A were proposed to be
associated with FRB-like radio transients by \citet{ban12}.  Neither
GRB has a redshift measurement, however, with distances estimated
through the empirical \citet{ama02} relation.  Moreover, further searches
for FRB-like radio emission associated with GRBs, and a joint
statistical analysis of a range of such experiments by \citet{pal14},
show that the FRB-like signals reported by \citet{ban12} are unlikely
to be astrophysical.  Thus, we do not consider these further as
appropriate for deriving EEP limits.  Conversely, the GRB-only
constraints from \citet{gao15} are still valid.

Clearly, obtaining accurate distances to FRBs is one key advance
required to utilise FRBs in the type of analysis proposed by
\citet{wei15}.  Very recently the first accurate localisation of an
FRB has led to the likely identification of an FRB host galaxy and the
measurement of an FRB redshift \citep{kea16}.  However, see
\citet{will16} for an alternate interpretation of the results
presented in \citet{kea16}.

\section{THE EEP CONSTRAINT FROM FRB 150418}
FRB 150418 was detected at the Parkes radio telescope on 2015 April 18
at 04:29:07.056\,UT at a frequency of 1382\,MHz, reported by
\citet{kea16}.  Rapid multi-wavelength follow-up revealed a fading
radio source at J2000 coordinates RA=$07^{\rm h}16^{\rm m}34\fs55$,
Dec$=-19\degr00\arcmin39\farcs9$ which \citet{kea16} argue is an afterglow
associated with the FRB.  At the position of the afterglow,
imaging on the Subaru and Palomar 200-inch telescope and spectroscopy
with Subaru and Keck reveal a $z=0.492\pm0.008$ galaxy, giving a
luminosity distance of approximately $d=2.5$ Gpc (for an assumed
cosmology of $H_{0}=69.6\,{\rm km\,s}^{-1}\,{\rm Mpc}^{-1}$,
$\Omega_{m}=0.286$, and $\Omega_{\Lambda}=0.714$; \citealt{wri06}).

From the celestial coordinates of FRB 150418, and taking the distance
to the centre of the Milky Way from Earth to be 8 kpc, the impact
parameter is $b=6.4$\,kpc.  The measured time delay, $\tobs$, for FRB
150418 was approximately 0.8\,s, between frequencies of approximately
1.2 and 1.5\,GHz.  Using the same value for $M_{\rm MW}=6 \times
10^{11}\,M_\odot$ as used by \citet{wei15} and their assumption that
$\tobs \gg \tdm$, the limit on $\Delta \gamma$ for FRB 150418 is
$\Delta \gamma<2.2\times10^{-8}$, for photons of frequency 1.2\,GHz
and 1.5\,GHz.

This limit is comparable to that inferred for FRB 110220 by
\citet{wei15}.  However, after accounting for dispersion due to the
host galaxy, the Milky Way, and the Milky Way halo, \citet{kea16}
estimate that the remaining dispersion due to the intergalactic medium
(IGM) is $540\pm140\,\dm$, which gives $\omegaigm=0.049\pm0.013$ from
the FRB 150418 data.  This is compared to expectations from
$\Lambda$CDM models and WMAP observations, which predict
$\omegaigm=0.041\pm0.002$ \citep{hin13}.  Both estimates agree to
within the errors, leading to the reasonable conclusion that \tobs\ is
dominated by dispersion and that the assumption adopted by
\citet{wei15} ---  $\tobs\gg\tdm$ --- is not needed for FRB 150418.

It is, however, still difficult in the case of FRB 150418 to determine
what fraction of \tobs\ could be attributable to gravitational delays.
Within the errors, \tobs\ for FRB 150418 is plausibly explained by
dispersion in and around the Milky Way, in the host galaxy, and in the
IGM.  However, the errors in the various DM components are of order
20\%.  For example, an error of 20\% is quoted by \citet{kea16} on the
largest non-IGM component of ${\rm DM}_{\rm MW}=189\,\dm$, equating to
$\sim$5\% of the total DM.  Also, a 100\,\dm error is quoted on the
IGM contribution to the DM, equating to $\sim$10\% of the total DM.
If errors in these quantities of 5\%--10\% magnitude masked
gravitational delays, the limit on $\Delta \gamma$ above would be
reduced by factors of 10--20, giving $\Delta \gamma<1.1\times10^{-9}$
(for a factor of 20).  This limit is more than an order of magnitude
lower than that obtained by \citet{wei15}.  We plot this limit along
with other differential limits on $\gamma$ in Figure~\ref{fig:gamma}.

\section{DISCUSSION}
With confirmation by \citet{kea16} that FRBs originate at cosmological
distances, they are now serious contenders as cosmological probes and
of great interest for their intrinsic astrophysics.  In addition to
the fundamental advance provided by \citet{kea16}, other recent
results are starting to reveal information on their progenitors and
environments \citep{mas15}.  As suggested by \citet{wei15}, future instruments such as the Square Kilometre Array (SKA) will be well-placed to open up new lines of investigation into astrophysics and cosmology using FRBs \citep{mac15}.  

In the specific use of FRBs to place limits on Einstein's Equivalence
Principle, the work of \citet{wei15} suggests great promise.  With a
distance now known for FRB 150418, improved limits are now possible
for this single FRB, likely approximately an order of magnitude better
than inferred by \citet{wei15}.  However, to achieve further
improvements in EEP limits using FRBs with the same technique will be
difficult.  The sensitivity of the limits to distance are not strong,
as shown by \citet{wei15} since, for cosmological FRBs, $d$ will
always be of order Gpc and $b$ will always be of order kpc, entering
as ln$(b/d)$ in the limit on $\Delta \gamma$.  Incremental
improvements for \frb\ may be possible through more realistic
gravitational potentials for the Milky Way \citep[e.g.,][]{bov15,wu16}, the
Local Group, and the host galaxy (once more mass estimates are
available).

However, substantial improvements are possible via other sources of
gravitational potential.  \citet{nus16} argue that the potential from
large-scale structure can lower the EEP limit by up to four orders of
magnitude.  This primarily comes from a statistical treatment of the
fluctuations in gravitational potential estimated from the cosmic
microwave background power spectrum (assuming linear structure
growth).  In a similar vein, \citet{zha16} estimate that EEP limits from FRBs could also be lowered by orders of magnitude if FRBs were identified behind galaxy clusters.

If the actual line-of-sight potential distribution
is known for a burst like \frb\ via optical  or \ion{H}{1} galaxy redshift
surveys (like \citealt{daw15} or \citealt{abd15}), a robust limit could
be determined for individual objects.  This would be in addition to
the type of improvement in $\Delta \gamma$ discussed above.

A significant part of the value of the known distance for FRB 150418
is in the confirmation that dispersion dominates \tobs.  The 
challenge in placing future limits on EEP from FRBs will be in
understanding the various dispersion components that contribute to
this dominance: the Milky Way (and halo) contribution; the
contribution from the FRB host galaxy; and the IGM contribution.
These contributions are all uncertain at the 10\% level.  These
uncertainties would need to be reduced to the $\sim1\%$ level to place
confident limits on differential gravitational delays and thus attain $\Delta
\gamma<10^{-10}$.

We can look to improve models of the Milky Way's ionized content.
This will likely come from a combination of FRBs that are nearby (on
the sky) to well-studied pulsars coupled with knowledge of the pulsar
distances \citep{del11,han15}. The SKA may be required to gather such
observations, in particular pulsar distances via parallax measurements
\citep{smi11,han15}.  Separately, searches for pulsars/FRBs in Local Group
galaxies \citep[e.g.,][]{bha11,rub13,rav15} could constrain the
dispersion measure of the halo.  If more than one source can be
detected in one of these galaxies (or even two cosmological FRBs in
the same host galaxy), they would have common IGM and Milky Way
(including halo) DM contributions but different host galaxy
contributions.  Such a scenario would provide some knowledge of the
magnitude (and variability) of the host galaxy DM contribution.

Recently, \citet{mas15} detected Faraday rotation and scintillation
for FRB 110523, providing evidence that this FRB occurred in a dense
magnetised plasma environment (also see \citealt{kul15}).  Assuming that
FRB 110523 originates in a host galaxy at cosmological distance,
analyses such as performed by \citet{mas15} may, in the future, help
provide some constraints on the DM contribution of the host galaxy.
This may start to address the challenges noted above.  

\begin{figure*}
%  \plottwo{deltagamma.pdf}{deltagamma_dE.pdf}
  \plottwo{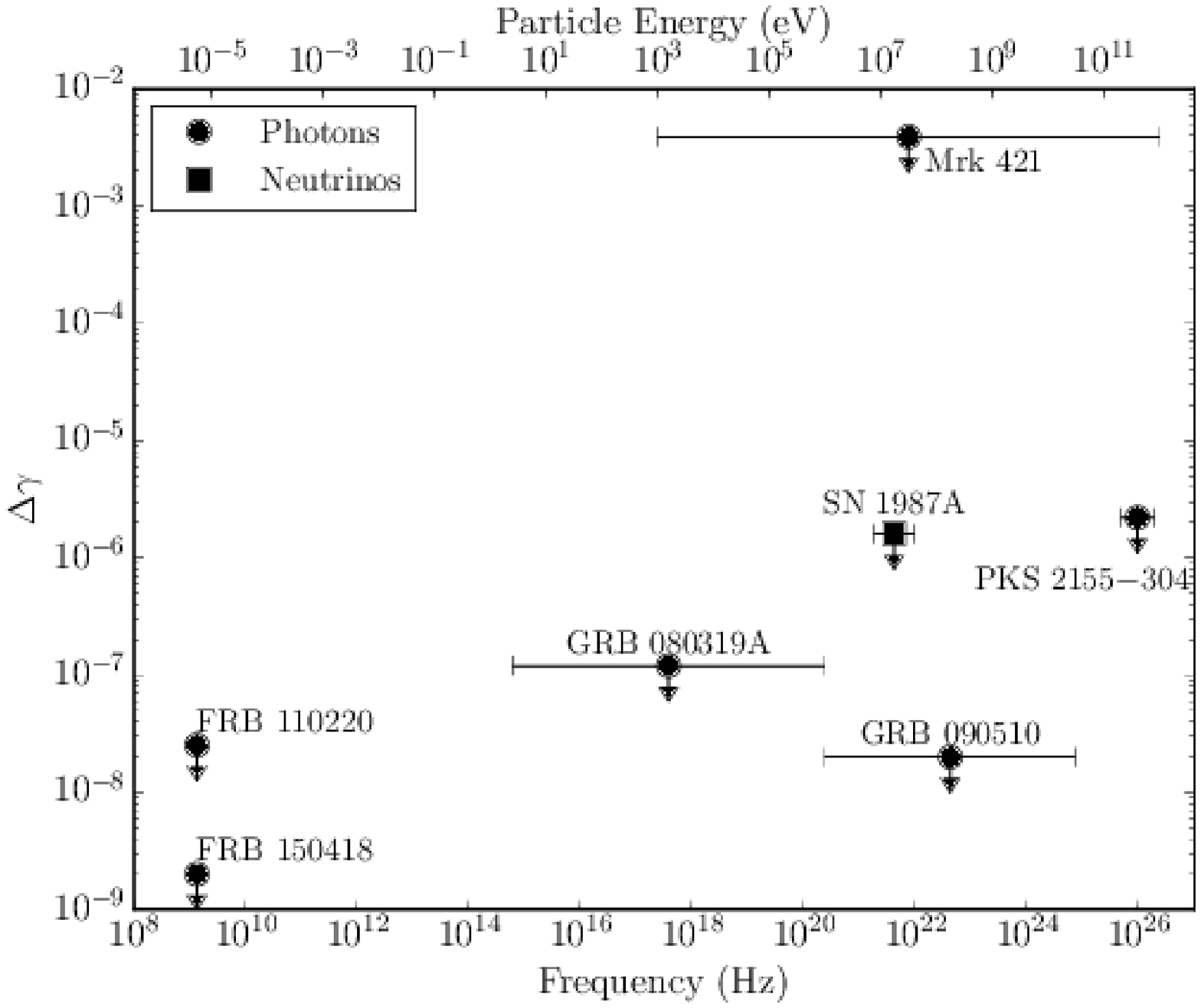}{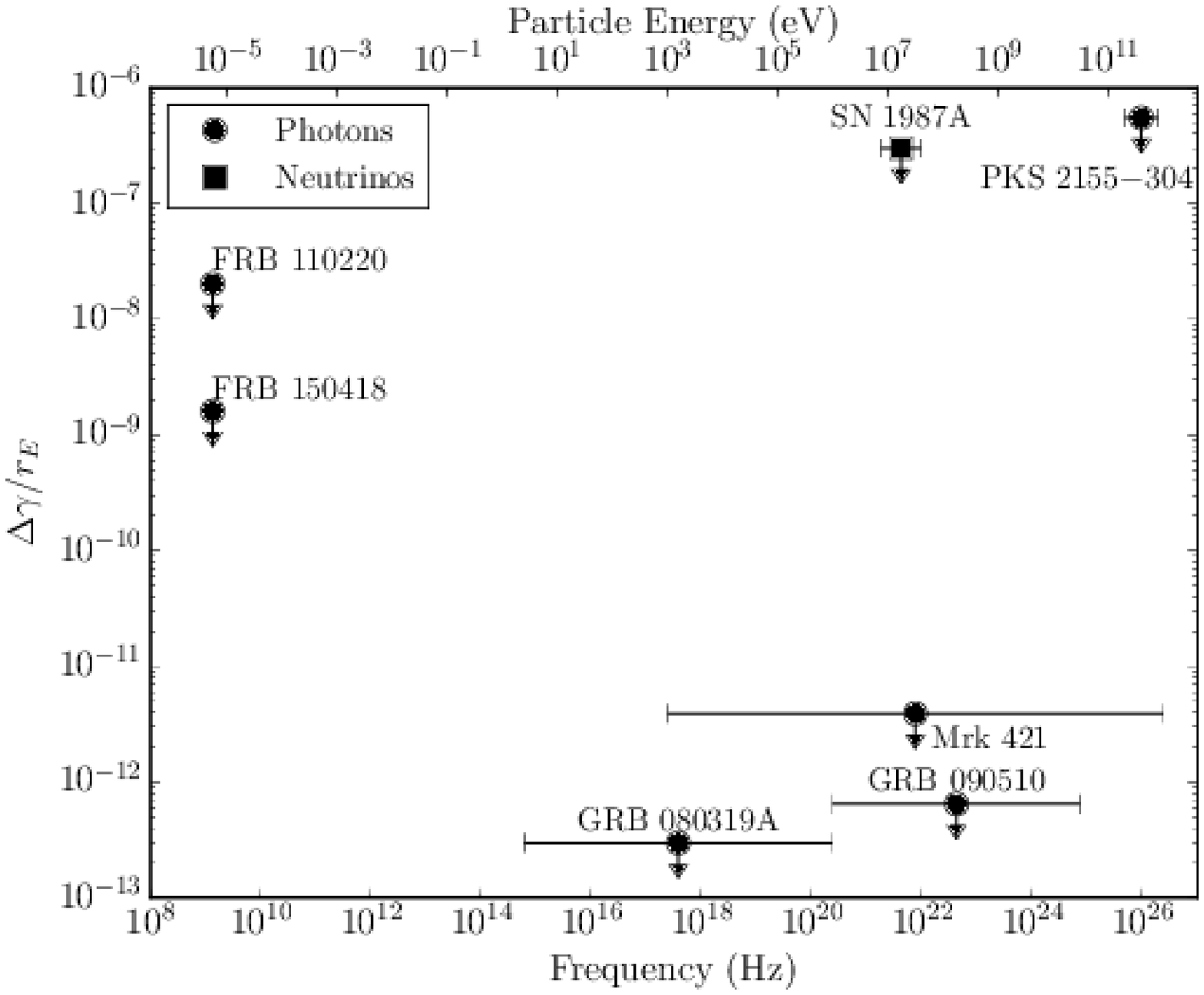}  
  \caption{Left: differential limits on the post-Newtonian parameter
    $\gamma$ versus particle energy or frequency, for two GRBs
    \citep{gao15}, SN~1987A \citep{lon88}, FRB~110220 \citep{wei15},
    the blazars Mrk~421 and PKS~2155$-$304 \citep{wei16},
    and FRB~150418 (this work).  The limit for SN~1987A is for
    neutrinos, while for the other sources the limit is for photons.
    In all cases the range of the error bars shows the range of
    particle energy used for the limit.  Right: differential limits on
    $\gamma$ divided by ratio of energies used $r_E\equiv E_{\rm
      hi}/E_{\rm lo}$ versus particle energy or frequency.}
  \label{fig:gamma}
\end{figure*}

The limits presented here are differential limits, $\Delta \gamma$.
Absent any particular form for EEP violation as a function of photon
energy, we consider how  our limits compare with those of
\citet{gao15} and others by looking at $\Delta \gamma/r_E$, where
$r_E=E_{\rm hi}/E_{\rm lo}$ is the ratio of high and low energies used
in the limit.  This is plotted in the right panel of
Figure~\ref{fig:gamma}, and as can be seen the multi-decade spread in
energies for GRBs and blazars \citep{wei16} makes their limits in this metric better than \frb\ by more than
three orders of magnitude.  However, this could potentially be rectified
by future improvements in FRB observing, as described above and
below.  Obtaining independent limits across a broad range of the
electromagnetic (EM) spectrum will be useful if considering any model of
EEP violation that has a frequence dependence, as well as studies of
the plasma dispersion and constraining progenitor models through
spectral indices \citep[e.g.,][]{tin15,kar15}.  The recent direct detection of gravitational waves now opens up possibilities for similar tests using non-EM channels.  For example, \citet{ell16} calculate an upper limit to the difference in the speed of gravitational waves and gamma-rays from the possible coincidence of the GW150914 detection and a gamma-ray transient reported by \citet{con16}.  Similarly, \citet{wan16} use the recent proposal that a gamma-ray outburst in PKS B1424$-$418 was associated with a PeV neutrino detection \citep{kad16}  to place limits on EEP.

While most FRBs detected thus far have
been between the frequencies of 1.2\,GHz and 1.5\,GHz at the Parkes
radio telescope, FRB 110523 was detected between 700\,MHz and 900\,MHz
with the Green Bank Telescope \citep{mas15}.  Simultaneous detections
at different telescopes may allow EEP limits over substantially larger
frequency ranges, improving $\Delta \gamma/r_E$ significantly. In
\citet{kea16} observations were done simultaneously with both Parkes
at 1.4\,GHz and the Murchison Widefield Array (MWA) at 150\,MHz,
although the latter was a non-detection.  Further FRB searches are
active at low frequencies ($\sim$100 MHz) with the MWA \citep{tin15} and LOFAR \citep{coe14,kar15} and at high
frequencies (10s of GHz) by the V-FASTR project at the VLBA
\citep{tro13}.  The larger frequency range also gives improved
measurements of Faraday rotation ($\propto \nu^{-2}$) and scattering
($\propto \nu^{-4}$) that can help separate local and IGM
contributions to time delay.  Extending this to searches at optical
and high-energy (X-ray, $\gamma$-ray) wavelengths could make this even
more powerful, although there is not yet any evidence of simultaneous
prompt emission from an FRB outside the radio band.

Further opportunities for testing the EEP could come from the
discovery of a multiply-imaged FRB lensed by a galaxy
(cf.\ \citealt{zhe14}), echoing the recent discovery of a
multiply-imaged supernova by \citet{kel15}\footnote{In the future, one
  could also conduct FRB search observations contemporaneous with
  searches for the late-appearing lensed component of multiply imaged
  supernovae \citep{kel15b}, to constrain progenitor models
  \citep[e.g.,][]{fal14,zha14}.}.  This would require interferometric
detection to achieve the needed spatial resolution, either with
contemporary inteferometers \citep{law15} or next-generation faclities
like the SKA.  In addition to the standard cosmological tests
\citep{ref64}, any deviations from the time delays expected purely
from lensing could limit not just differential acceleration, $\Delta
\gamma$, but the absolute acceleration $\gamma$.  However, this relies
on the precision of the lensing models, which in the case of
SN~Refsdal is limited because of the massive galaxy as well as a
surrounding galaxy cluster. There could also be differences in
dispersive delays through the cluster, but for typical intracluster
medium densities of $\sim 10^{-3}\,{\rm cm}^{-3}$ and lensing path
delays of $\sim 10\,$d, the difference in DM would be only
$10^{-5}\,\dm$.  Searches for FRBs toward massive galaxy clusters may
be worthwhile even without the detection of of multiply-imaged bursts,
as the cluster would give an additional DM contribution of $\sim
10^3\,\dm$ (for a cluster size of 1\,Mpc) which could be combined with
measurements of the X-ray emission and the Sunyaev-Zeldovich (SZ)
effect \citep{sun80} to strongly constrain the hot gas in the cluster
and aid in SZ cosmology.  The polarisation observed for FRBs may also
be put to use in experiments utilizing a lensed FRB \citep{pre02}.

%$z$ from \ion{H}{1}

%% \begin{deluxetable*}{l c c c c}
%% \tablewidth{0pt}
%% \tablecaption{Contributions and Uncertainties to the Dispersion and
%%   Time Delay of \frb\label{tab:eep}}
%% \tablehead{
%% \colhead{Component} & \colhead{DM} & \colhead{DM Uncertainty} &
%% \colhead{Cumulative DM} & 
%% \colhead{Cumulative Delay}\\
%% & \colhead{(\dm)} & \colhead{(\dm)} & \colhead{(ms)}
%% }
%% \startdata
%%   Milky Way & 189 & \phn37 & $189\pm 37$\phn &$208\pm41$\phn\\
%%   Milky Way Halo & \phn30 & \nodata &$219\pm37$\phn & $241\pm41$\phn\\
%%   IGM\tablenotemark{a} & 445 & 102 &$664\pm108$ &$732\pm120$\\
%%   Host Galaxy\tablenotemark{b}& \phn17 & \nodata &$681\pm108$ & $751\pm120$ \\
%% \hline
%%   Measured  & & & 776 & 855 \\
%%   \hline
%%   Residual & & & $95\pm108$ & $104\pm120$
%%   \enddata
%% \tablecomments{All components and uncertainties are taken from
%%   \citet{kea15}, unless otherwise noted.}
%%   \tablenotetext{a}{Estimated from $\Omega_{\rm b}=0.046\pm0.002$ \citep{hin13},
%%   assuming $\Omega_{\rm IGM}=0.9 \Omega_{\rm b}$ with an additional
%%   $100\,\dm$ uncertainty.}
%%   \tablenotetext{b}{Effective DM, divided by $(1+z)^2$.}
%% \end{deluxetable*}

\section{Acknowledgements}
We thank an anonymous referee for helpful corrections.  SJT is a
Western Australian Premier's Research Fellow.  DLK is supported by NSF
grant AST-1412421.

\bibliographystyle{apj} 
%\bibliography{frb}

\end{document}